\documentclass[draftcls,letterpaper,onecolumn, 12pt]{IEEEtran}

\begin{document}
%
\title{A Family of Binary Sequences with Optimal Correlation Property and Large Linear Span}
\date{August 22, 2005}
%
%
\author{Xiangyong~Zeng,
        Lei~Hu, Qingchong~Liu, {\em Member}
\thanks{X. Zeng and Q. Liu's work is supported in part by ARO under Grant W911NF-04-1-0267 and NSF under Grant
CNS-0435341. L. Hu's work is supported in part by the National
Science Foundation of China (NSFC) under Grants No.60373041 and
No.90104034.}
\thanks{X. Zeng is with Department of Electrical and System
Engineering, Oakland University, Rochester, MI 48309, USA and with
the Faculty of Mathematics and Computer Science, Hubei University,
Xueyuan Road 11, Wuhan 430062, P. R. China. Email:
zeng2@oakland.edu, xzeng@hubu.edu.cn}
\thanks{L. Hu is with the Graduate School of
Chinese Academy of Sciences, 19A Yuquan Road, Beijing 100049, P.
R. China.}
\thanks{Q. Liu is with Department of Electrical and System
Engineering, Oakland University, Rochester, MI 48309, USA.}}
\maketitle

\begin{abstract} A family of binary sequences is presented and
proved to have optimal correlation property and large linear span.
It includes the small set of Kasami sequences, No sequence set and
TN sequence set as special cases. An explicit lower bound
expression on the linear span of sequences in the family is given.
With suitable choices of parameters, it is proved that the family
has exponentially larger linear spans than both No sequences and
TN sequences. A class of ideal autocorrelation sequences is also
constructed and proved to have large linear span.
\end{abstract}


%
\IEEEpeerreviewmaketitle

\clearpage

\section{Introduction}
%
%
%
%

Binary sequences are important for CDMA systems, spread spectrum
systems, and broadband satellite communications \cite{1}. Families
of sequences for such applications are desired to have low
autocorrelation, low cross-correlation, and large linear span
\cite{2,3}. Families of Gold-pairs \cite{4,5} and bent function
sequences \cite{6,7} as well as the small and large families of
Kasami sequences \cite{8,9} all have desirable correlation
properties. However, these sequences, except the bent function
functions, have small values of linear span. In \cite{10},
relaxing correlation, Gong constructed a sequence set with a much
larger linear span.

\vspace{2mm}

Important results are obtained for increasing linear span of
sequences while keeping the sequences optimal in correlation
respect to the Welch bound \cite{11,12,13}. No and Kumar \cite{12}
proposed a family of No sequences of period $2^n-1$, which are
defined by
\begin{equation}
s_h(t)=tr^m_1\{[tr^n_m(\alpha^{2t})+\gamma_h\alpha^{(2^m+1)t}]^r\}
\end{equation}
where $n=2m$, $\alpha$ is a primitive element of the finite field
$F_{2^n}$, $\gamma_h$ ranges over each element of $F_{2^{n/2}}$
exactly once as $h$ ranges from $1$ to $2^{n/2}$, and $r$ is an
integer with $1\leq r<2^m-1$ such that ${\rm gcd}(r,2^m-1)=1$.
When $r=1$, the family is the small set of Kasami sequences. The
maximal linear span of No sequences is $O(n\cdot
4^{\frac{n}{4}})$. Klapper \cite{13} generalized the family to
that of so-called Trace-Norm (TN) sequences with an expression of
the form
\begin{equation}
s_h(t)=tr^{m}_{1}\{[tr^{mk}_m(tr^n_{mk}(\alpha^{2t})+\gamma_h\alpha^{(2^{mk}+1)t})]^r\}
\end{equation}
where $n,\alpha, \gamma_h$ and $r$ are the same as for the
No-Kumar family, while other two parameters $m$ and $k$ satisfy
$mk=\frac{n}{2}$. For suitable $k$ and $r$, TN sequences have much
larger linear spans than that of No sequences, and their maximal
linear span is $O(n\cdot 5^{\frac{n}{4}})$. These constructions
were extended by No {\it et al.} in 1997 \cite{14}, and
generalized by Gong in 2002 \cite{10}. These extended and
generalized families of sequences have the same correlation
properties as that of No sequences and TN sequences \cite{10}. It
remains unanswered whether a family of sequences with larger
linear span exists.

\begin{table*}\caption{Families of binary sequences of period $2^n-1$
with optimal correlation $R_{\rm max}=2^{\frac{n}{2}}+1$}
\begin{center}
  \begin{tabular*}{0.72\textwidth}
     {@{\extracolsep{\fill}}ccccc}
     \hline  \hline\\
 Family & $n$&Family size&Maximum linear span\\[0.7ex]
 \hline
  Bent function sequences&$4m$&$2^{\frac{n}{2}}$&$\geq\left({n/2}\atop {n/4}\right)2^{{n/2}}$ \\[0.7ex]

 Small set of Kasami sequences&2m&$2^{\frac{n}{2}}$&$\frac{3n}{2}$\\[0.7ex]

No sequences&2m&$2^{\frac{n}{2}}$&$n(2^{\frac{n}{2}}-1)/2$\\[0.7ex]

TN sequences&2mk&$2^{\frac{n}{2}}$&$>3n(3k-1)^{m-2}/2$\\[0.7ex]

Sequences we studied&2mk&$2^{\frac{n}{2}}$&$>3^{k-1}n[2^{k-2}(3k-1)]^{m-2}/2$\\[0.7ex]
\hline

    \end{tabular*}
  \end{center}
\end{table*}

In this correspondence, we study the linear span of binary
sequences defined by
\begin{equation}
 s_h(t)=\sum_{i\in I}\{tr^{mk}_m[(tr^n_{mk}(\alpha^{2t})+\gamma_h\alpha^{(2^{mk}+1)t})^u]\}^i
\end{equation}
where $n,m,k,\alpha, \gamma_h$ and $r$ are the same as for the
Klapper family, and $u$ satisfies $1\leq u<2^{mk}-1$ and ${\rm
gcd}(u,2^{mk}-1)=1$. The index set $I$ is chosen such that for a
primitive element $\beta$ of $F_{2^m}$, $\{\sum\limits_{i\in
I}\beta^{it_1}\}_{t_1=0}^{\infty}$ represents an ideal
autocorrelation sequence of period $2^m-1$. When
$I=\{r,2r,\cdots,2^{m-1}r\}$ and $u=1$, sequences in Eq. (3) is
the Klapper sequences.

\vspace{2mm}

This family of sequences defined by Eq. (3) can be regarded as a
special case of the generalized Kasami signal set \cite{10}, whose
linear span was not considered. This correspondence gives a lower
bound on linear spans of sequences in Eq. (3). More precisely, for
$u=\sum\limits_{j=0}^{k-2}2^{mj}$ and $3\leq k \leq 5$, we prove
that a majority of sequences in the family have linear spans at
least $O(n\cdot 2^{\frac{2n}{3}})$
($2^{\frac{2n}{3}}>6.32^{\frac{n}{4}}$), which is significantly
larger than the linear span $O(n\cdot 4^{\frac{n}{4}})$ for
No-Kumar sequences and $O(n\cdot 5^{\frac{n}{4}})$ for Klapper
sequences. Table I summarizes the family size and linear span
properties of the families mentioned above.

\vspace{2mm}

The family of sequences in Eq. (3) has optimal correlation
property. It contains an ideal autocorrelation sequence \cite
{15}, whose linear span, although very large, is much less than
that of any other sequence in the family. In order to obtain
sequences with ideal autocorrelation and large linear span, we
will further specify $I$ to relate it to Legendre sequences and
derive a tighter lower bound on the linear span of this ideal
autocorrelation sequence. This will be presented in Section IV.

\vspace{2mm}

The remainder of this correspondence is organized as follows.
Section II gives some necessary notations and preliminary lemmas.
Section III derives lower bounds on linear spans of the sequences.
Section IV shows that a class of sequences with ideal
autocorrelation property also have large linear span. Section V
concludes the study.

\vspace{3mm}

\section{Preliminaries}

Let $\mathcal{F}$ be the family of $M$ binary sequences of period
$N=2^n-1$ given by
$$
\mathcal{F}=\{\{s_h(t),\,0\leq t\leq N-1\}\,\,|\,0\leq h\leq
M-1\}.
$$
The {\it correlation function} of the sequences $\{s_h(t)\}$ and
$\{s_l(t)\}$ in $\mathcal{F}$ is
$$
 R_{h,l}(\tau)=\sum\limits^{N-1}_{t=0}(-1)^{s_h(t)-s_l(t+\tau)}
$$
where $0\leq h, l\leq M-1$, and $0\leq \tau\leq N-1$. The {\it
maximum magnitude} $R_{\rm max}$ of the correlation values is
$$
 R_{\rm max}={\rm max} \; |R_{h,l}(\tau)|
$$
where $ 0\leq h, l\leq M-1$, $ 0\leq \tau\leq N-1$, and the cases
of in-phase autocorrelations ($h=l\,\,{\rm and}\,\,\tau=0$) are
excluded. A family of binary sequences of period $2^n-1$ is said
to have {\it optimal correlation property} if $R_{\rm max}\leq
2^{\frac{n}{2}}+1$.  For  $h=l$, $R_{h,l}(\tau)$, abbreviated by
$R_h(\tau)$, is the {\it autocorrelation function} of
$\{s_h(t)\}$. The sequence $\{s_h(t)\}$ is said to have an {\it
ideal autocorrelation property} if
$$
R_h(\tau)=\left\{\begin{array}{rl}N&{\rm if}\, \tau\equiv 0\, {\rm
mod}\,N; \\-1&{\rm otherwise.}\end{array}\right.
$$

\vspace{2mm}

Let $F_{2^n}$ be the finite field with $2^n$ elements, and $n=em$
for some positive integers $e$ and $m$. The {\it trace function}
$tr^n_m(\cdot)$ from $F_{2^n}$ to $F_{2^m}$ is defined by
$$
tr^{n}_{m}(x)=\sum\limits^{e-1}_{i=0}x^{2^{im}}
$$
where $x$ is an element in $F_{2^n}$.

\vspace{2mm}

The trace function has the following properties \cite{16}:

\vspace{1mm}

i) $tr^n_m(ax+by)=a\cdot tr^n_m(x)+b\cdot tr^n_m(y)$, for all $a$,
$b$ $\in$ $F_{2^m}$, $x$, $y$ $\in$ $F_{2^n}$.

ii) $tr^n_m(x^{2^m})=tr^n_m(x)$, for all $x$ $\in$ $F_{2^n}$.

\vspace{1mm}

iii) $tr^n_1(x)=tr^m_1(tr^n_m(x))$, for all $x$ $\in$ $F_{2^n}$.

\vspace{3mm}

The operation of multiplying by $2$ divides the integers modulo
$2^m-1$ into sets called the {\it cyclotomic cosets} modulo
$2^m-1$. The cyclotomic coset containing $s$ is $\{s$, $2s$,
$2^2s$, $\cdots$, $2^{e_s-1}s\}$, where $e_s$ is the smallest
positive integer such that $2^{e_s}s\equiv s$ (mod $2^m-1$).
Furthermore, $e_s$ divides $m$, and $e_s=m$ for $m$ prime and
$s\not\equiv 0\,(\mathrm{mod}2^m-1)$. The smallest positive
integer in the cyclotomic coset $\{s,2s,2^2s,\cdots,2^{e_s-1}s\}$
is called its {\it coset leader} \cite{17}.

\vspace{3mm}

For two integers $a$ and $b$ with $a\leq b$, let $[a,b]$ be the
interval consisting of all integer $c$ with $a\leq c\leq b$, and
the {\it length} is $b-a+1$. When $a=b$, $[a,b]$ is called a {\it
single point interval} and written as $[a]$. Two intervals $[a,b]$
and $[c,d]$ are {\it un-incorporative} if $b+2\leq c$ or $d+2\leq
a$. A set of several pairwisely un-incorporative non-negative
intervals $\{[a_j,b_j] \,|\,j\in J\}$ determines a positive
integer $\sum\limits_{j\in J}\sum\limits_{x\in [a_j,b_j]}2^{x}$,
where $J$ is an index set. For a positive integer $c$, there exist
an index set $K$ consisting of non-negative integers such that
$c=\sum\limits^{}_{k\in K}2^k,$ which determines a set of
un-incorporative intervals $\{[a_j,b_j] \,|\,j\in J\}$ such that
$\bigcup_{j\in J}[a_j,b_j]=K$. This fact will be used in
derivation of the main result in this correspondence.

\vspace{3mm}

The following notations are used in the rest of this
correspondence:

\begin{itemize}

\item $m$, $k$, and $n$: positive integers, $n=2mk$;

\item $N=2^n-1$, $M=2^m-1$, and
$T=\frac{N}{M}=\frac{2^n-1}{2^m-1}$;

\item $F_{2^n}$: the finite field with $2^n$ elements;

\item $\alpha$: a primitive element of $F_{2^n}$;

\item $\beta=\alpha^T$: a primitive element of $F_{2^m}$;

\item $\Gamma(m)$: the set consisting of all non-zero coset
leaders modulo $2^m-1$;

\item $C_i=\{i2^j\,({\rm mod}\,2^m-1)\,|\,j=0,1,\cdots,m\}$, i.e.,
the cyclotomic coset modulo $2^m-1$ containing the element $i$;

\item $e_i=|C_i|$;

\item $Z_p$: a residue ring of integers modulo $p$;

\item $V=\{0,1,\cdots,k-1\}\,\}$, where $k$ is a positive integer;

\item $V^t=V\times V\times\cdots\times V$ is the Cartesian product
of $t$ copies of $V$;

\item $w(i)$: the weight of integer $i$, i.e., the number of ones
in the coefficients of the binary expansion of $i$;

\item $\lfloor z\rfloor$: the largest integer not exceeding $z$;

\item $[a,b]$: the integer interval consisting of all integer $c$
with $a\leq c\leq b$.

\item $\gamma_0=0,\gamma_1,\cdots,\gamma_{2^{mk}-1}$: all $2^{mk}$
elements of the field $F_{2^{mk}}$.

\end{itemize}

\vspace{3mm}

The following lemmas will be used to prove our results.

\vspace{1mm}

{\it Lemma 1} (Proposition 1, \cite{10}, or Theorem 5, \cite{15}):
Let $I\subseteq Z_M$ be an index set. If the binary sequences
$\{a(t_1)\}$ of period $M$ given by
\begin{equation}
a(t_1)=\sum_{i\in I}\beta^{it_1}  \label{eqat1}
\end{equation} has the ideal autocorrelation
property, then so does the binary sequence $c(t)$ of period $N$
defined by
\begin{equation}
c(t)=\sum_{i\in I}\{tr^{mk}_m[(tr^n_{mk}(\alpha^{2t}))^u]\}^{i}
\end{equation}
for any $u$ satisfying ${\rm gcd}(u,2^{mk}-1)=1$.

\vspace{2mm}

It is noted that any binary sequence of period $2^m-1$ with ideal
autocorrelation property can be written as
\begin{equation}
a(t_1)=\sum_{0\leq i\leq 2^m-2}A_i\beta^{it_1}
\end{equation}
where $A_i\in \{0,1\}$ and $A_i$ take a same value on each
cyclotomic coset modulo $2^m-1$, i.e., $A_{2i}=A_{i}$ for any
$1\leq i<2^{m}-1$ \cite{18}.

\vspace{2mm}

Take $I$ as the set of all $i$ with $A_i\not=0$, then Eq. (6)
becomes Eq. (4). $I$ is a union of several cyclotomic cosets, and
$I=\cup_{i\in I\cap \Gamma(m)}C_{i}$. For the goal of obtaining a
good lower bound on linear span, we assume in this correspondence
that there is an $i_0\in I$ such that gcd$(i_0,2^m-1)=1$ (This is
an assumption satisfied by many ideal autocorrelation sequences).
Since gcd$(2^{m-1}-1,2^m-1)=1$, replacing $\beta$ with
$\beta^{(2^{m-1}-1)i_0^{-1}\,{\rm mod}\,2^m-1}$, we can further
assume $2^{m-1}-1\in I$.

\vspace{3mm}

The optimal correlation property of the family of sequences
expressed in Eq. (7) is restated as Lemma 2 for completeness.

\vspace{1mm}

{\it Lemma 2:} (Proposition 2, \cite{10}) Let
$\gamma_0=0,\gamma_1,\cdots,\gamma_{2^{mk}-2}$, and
$\gamma_{2^{mk}-1}$ be the all elements of $F_{2^{mk}}$. For
$0\leq h\leq 2^{mk}-1$, define $\{s_h(t)\}$ as the sequence by
\begin{equation}
s_h(t)=\sum_{i\in
I}\{tr^{mk}_m[(tr^n_{mk}(\alpha^{2t})+\gamma_h\alpha^{(2^{mk}+1)t})^u]\}^{i}
\end{equation}
where $I$ is the index set as mentioned above, and $1\leq u\leq
2^{mk}-1$ is an integer relatively prime to $2^{mk}-1$. Then the
family $\mathcal{F}$
\begin{equation}
\mathcal{F}=\{\{s_h(t)\}_{0\leq t< 2^n-1}\,|\,0\leq h\leq
2^{mk}-1\} \label{eqF}
\end{equation}
of $2^{mk}$ binary sequences of period $N$ is an optimal
correlation sequence set with respect to Welch's bound.
Furthermore, $R_{h,k}(\tau) \in \{-1, 2^{\frac{n}{2}}-1,
-2^{\frac{n}{2}}-1\}$ for any out-of-phase shift $(h,k,\tau)$
($h\not=k$ or $\tau\not=0$).

By Lemma 1, $\{s_0(t)\}$ is an ideal autocorrelation sequence.

\section{Linear span of sequences}

This section proves sequences in the family $\mathcal{F}$ have
large linear span.

The linear span of a sequence is the smallest degree of which a
linear recursion satisfied by the sequence exists. Key \cite{20}
described a method for determining the linear span of a binary
sequence of period $2^n-1$. The linear span of $\{s_h(t)\}_{0\leq
t< 2^n-1}$ can be determined by expanding the expression of
$s_h(t)$ as a polynomial in $\alpha^t$ of degree less than $2^n-1$
and then counting the number of monomials in $\alpha^t$ with
nonzero coefficients occurring in the expansion. This technique
will be applied to determine the linear span of sequences in
family $\mathcal{F}$.

\vspace{2mm}

Denote each exponent $i\in I$ in Eq. (5) as
\begin{equation}i=2^{i_1}+2^{i_2}+\cdots+2^{i_{w(i)}}\end{equation} where
$0\leq i_1<i_2<\cdots<i_{w(i)}\leq m-1$.

Let $x=\alpha^{t}$ and $y=x^{2^{mk}-1}$. Substituting Eq. (9) into
Eq. (7). Then $s_h(t)$ can be written as
\begin{equation}\begin{array}{rcl}s_h(t)&=&\sum\limits_{i\in
I}[\sum\limits^{k-1}_{v=0}(\alpha^{2t}+\gamma_h\alpha^{(2^{mk}+1)t}+\alpha^{2^{mk+1}t})^{u2^{mv}}]^{i}
\\&=&\sum\limits_{i\in
I}(\sum\limits^{k-1}_{v=0}[x^2(1+\gamma_h
y+y^{2})]^{u\cdot2^{mv}})^{i}\\&=&\sum\limits_{i\in
I}\prod\limits^{w(i)}_{j=1} \sum\limits^{k-1}_{v=0}[x^2(1+\gamma_h
y+y^{2})]^{u\cdot2^{mv+i_j}}\\&=&\sum\limits_{i\in
I}\sum\limits_{\underline{v}\in V^{w(i)}}[x^2(1+\gamma_h
y+y^2)]^{\delta(i,\underline{v})}\end{array}\end{equation} where
$V=\{0,1,\cdots,k-1\}$,
$\underline{v}=(v_1,v_2,\cdots,v_{w(i)})\in V^{w(i)}$, and
\begin{equation}
\delta(i,\underline{v})=\sum\limits_{j=1}^{w(i)}u\cdot2^{mv_j+i_j}.\end{equation}

\vspace{3mm}

As the first step to count the number of monomials in $\alpha^t$
with nonzero coefficients occurring in right side of Eq. (10), we
show the following

\vspace{2mm}

{\it Lemma 3:} For different pairs $(i,\underline{v})$ and
$(i',\underline{v'})$, there is no monomial that appears with
nonzero coefficients in the expansions of both $(x^{2}(1+\gamma
y+y^{2}))^{\delta(i,\underline{v})}$ and $(x^{2}(1+\gamma
y+y^{2}))^{\delta(i',\underline{v'})}$.

\vspace{2mm}

{\it Proof:} Since $y=x^{2^{mk}-1}$, each monomial in $x$ in the
expansion of $x^{2}(1+\gamma_h y+y^{2})$ has an exponent (respect
to $x$) congruent to 2 modulo $2^{mk}-1$. Thus, each monomial in
the expansion of $(x^{2}(1+\gamma_h
y+y^{2}))^{\delta(i,\underline{v})}$ has an exponent congruent to
$2\cdot \delta(i,\underline{v})$ modulo $2^{mk}-1$.

If there is a monomial that appears with nonzero coefficients in
the expansions of both $(x^{2}(1+\gamma
y+y^{2}))^{\delta(i,\underline{v})}$ and $(x^{2}(1+\gamma
y+y^{2}))^{\delta(i',\underline{v'})}$, then
\begin{equation}2\cdot \delta(i,\underline{v})\equiv 2\cdot
\delta(i',\underline{v'})\,{\rm mod}\,(2^{mk}-1).\end{equation}
The integer $2u$ is relatively prime to $2^{mk}-1$. By Eq. (11)
and Eq. (12), we have
\begin{equation}\sum\limits_{j=1}^{w(i)}2^{mv_j+i_j}\equiv
\sum\limits_{j=1}^{w(i')}2^{mv'_j+i'_j}\,{\rm
mod}\,(2^{mk}-1).\end{equation}
Notice that
$$\sum\limits_{j=1}^{w(i)}2^{mv_j+i_j}\leq
\sum\limits_{j=1}^{w(i)}2^{m(k-1)+i_j}=2^{m(k-1)}i<2^{mk}-1.$$
Similarly, $\sum\limits_{j=1}^{w(i')}2^{mv'_j+i'_j}<2^{mk}-1$. Eq.
(13) can be written as
\begin{equation}\sum\limits_{j=1}^{w(i)}2^{mv_j+i_j}=\sum\limits_{j=1}^{w(i')}2^{mv'_j+i'_j}.\end{equation}
Since $2^{mv_j}\equiv 1\,{\rm mod\, (2^m-1)}$ and
$2^{mv_j+i_j}\equiv 2^{i_j}\,{\rm mod\, (2^m-1)}$, by Eq. (14),
one has
\begin{equation}\sum\limits_{j=1}^{w(i)}2^{i_j}\equiv\sum\limits_{j=1}^{w(i')}2^{i'_j}\,{\rm
mod\, (2^m-1)}.\end{equation} Since the both sides of Eq. (15) are
less than $2^m-1$, then
$$i=\sum\limits_{j=1}^{w(i)}2^{i_j}=\sum\limits_{j=1}^{w(i')}2^{i'_j}=i'.$$
Eq. (14) can be written as
\begin{equation}\sum\limits_{j=1}^{w(i)}2^{mv_j+i_j}=\sum\limits_{j=1}^{w(i)}2^{mv'_j+i_j}.\end{equation}
Since $0\leq i_1<i_2<\cdots<i_{w(i)}\leq m-1$, $mv_j+i_j$ are
pairwise incongruent modulo $m$ for all different $j$. This
implies that the two sides of Eq. (16) are the binary expansions
of the same integer, and hence,
$$\{mv'_j+i_j:
1\leq j\leq w(i)\}=\{mv'_j+i_j: 1\leq j\leq w(i)\}.$$ Comparing
the integers with the same remainder modulo $m$, we have
$v_j=v'_j$ for all $j$, i.e., $\underline{v}=\underline{v'}$.
Thus, $(i,\underline{v})=(i',\underline{v'})$. The proof ends.

\vspace{3mm}

Let $\rho (i,\underline{v})$ denote the number of monomials in $y$
appearing in the expansion of $(1+\gamma_h
y+y^2)^{\delta(i,\underline{v})}$ with nonzero coefficients. By
Eq. (10) and Lemma 3, we have
\begin{equation}LS(\{s_h(t)\})=\sum\limits_{i\in I}\sum\limits_{\underline{v}\in
V^{w(i)}}\rho (i,\underline{v}).\end{equation} Furthermore, Eq.
(17) can be written as follows.

\vspace{2mm}

{\it Proposition 4:}
\begin{equation}LS(\{s_h(t)\})=\sum\limits_{i\in I\cap \Gamma(m)}\sum\limits_{\underline{v}\in
V^{w(i)}}e_i\cdot \rho (i,\underline{v}).\end{equation}

\vspace{2mm}

{\it Proof:} Note that $I$ is a union of several cyclotomic
cosets, i.e., $I=\cup_{i\in I\cap \Gamma(m)}C_{i}$, to prove Eq.
(18), it is sufficient to show
\begin{equation}\sum\limits_{\underline{v}\in V^{w(i)}}\rho
(i,\underline{v})=\sum\limits_{\underline{v}\in V^{w(i')}}\rho
(i',\underline{v})\end{equation} holds for any $i,i'\in I$ with
$i\equiv 2i'\,\,\mathrm{mod}(2^m-1)$.

In Eq. (10), let
$$\Delta (x)=\sum\limits^{k-1}_{v=0}(x^{2}+\gamma_hx^{(2^{mk}+1)}+x^{2^{mk+1}})^{u2^{mv}}
=tr^{mk}_m[(x^2(1+\gamma_hy+y^2))^u].$$ For any $x\in F_{2^n}$,
$\Delta (x)\in F_{2^m}$ and hence $\Delta (x)^{i}=(\Delta
(x)^{i'})^2$ if $i\equiv 2i'\,\,\mathrm{mod}(2^m-1)$. From Eq.
(10), one has
$$\Delta(x)^{i}=\sum\limits_{\underline{v}\in V^{w(i)}}[x^2(1+\gamma_h
y+y^2)]^{\delta(i,\underline{v})},$$ and then
$$\sum\limits_{\underline{v}\in V^{w(i)}}[x^2(1+\gamma_h
y+y^2)]^{\delta(i,\underline{v})}=\{\sum\limits_{\underline{v}\in
V^{w(i')}}[x^2(1+\gamma_h
y+y^2)]^{\delta(i',\underline{v})}\}^2.$$ Since $(\Delta
(x)^{i'})^2$ and $\Delta (x)^{i'}$ have the same number of nonzero
monomials in their expansions, comparing the numbers of nonzero
monomials in the expansions of the both sides of the above
equality, Eq. (19) holds.

\hskip 3pt

By Proposition 4, the linear span can be determined by finding
$\rho (i,\underline{v})$ for all $i\in I\cap \Gamma(m)$ and
$\underline{v}\in V^{w(i)}$.

\vspace{2mm}

No and Kumar \cite{12} determined the number of nonzero monomials
in the expansion of $(1+\gamma_h y+y^2)^j$ for $j< 2^{mk}-1$. When
$j\geq 2^{mk}-1$, we can replace $j$ with
$j\,\,\mathrm{mod}(2^{mk}-1)$. Then, $\rho (i,\underline{v})$
equals to the number of nonzero monomials in the expansion of
$(1+\gamma_h y+y^2)^{\delta'(i,\underline{v})}$, where
$\delta'(i,\underline{v})$ is the remainder of
$\delta(i,\underline{v})$ modulo $2^{mk}-1$.

For $\gamma_h\neq 0$, define $\varepsilon_h=-1$ if the quadratic
$y^2+\gamma_h\cdot y+1=0$ is reducible over $F_{2^{mk}}$, and
$\varepsilon_h=1$ otherwise. Let $c_h$ be an integer with $0\leq
c_h\leq 2^{mk-1}$ such that
 \begin{equation}\delta_h=\left\{\begin{array}{ll}\alpha^{c_h(2^{mk}+1)}&{\rm if}\,\,\varepsilon_h=-1\\
 \alpha^{c_h(2^{mk}-1)}&{\rm if}\,\,\varepsilon_h=1\end{array}\right.\end{equation} is a root
of $y^2+\gamma_h\cdot y+1=0$. Let $g_h={\rm
gcd}(c_h,2^{mk}+\varepsilon_h)$. Then, $g_h<2^{mk-1}$ \cite{12}.

\vspace{2mm}

Let $R(i,\underline{v})$ be the total number of 1-runs occurring
within the binary expansion of $\delta'(i,\underline{v})$, and
$L(i,\underline{v},j)$ be the length of the $j$-th 1-run, $1\leq
j\leq R(i,\underline{v})$, with the runs being consecutively
numbered from the least to the most significant bits. Then,
$\delta'(i,\underline{v})$ can be written as
$$\delta'(i,\underline{v})=\sum^{R(i,\underline{v})}
_{j=1}2^{d_j}\cdot(\sum^{L(i,\underline{v},j)}_{l=0}2^l),$$ where
$d_j$ denotes the lowest exponent of 2 associated with the $j$-th
1-run.

\vspace{2mm}

By Theorem 2 in \cite{12}, the number of monomials with nonzero
coefficients appearing in the expansion of $(1+\gamma_h
y+y^{2})^{\delta'(i,\underline{v})}$ is
\begin{equation}\rho(i,\underline{v})=\prod\limits^{R(i,\underline{v})}
_{j=1}\{2^{L(i,\underline{v},\,j)+1}-1-2\lfloor
\frac{(2^{L(i,\underline{v},\,j)}-1)g_h}{2^{mk}+\varepsilon_h}
\rfloor\}.\end{equation}

When $\gamma_h=0$, one has
\begin{equation}\rho(i,\underline{v})=2^{\tau (i,\underline{v})}\end{equation} \cite{12},
where $\tau (i,\underline{v})$ is the weight of
$\delta'(i,\underline{v})$. It was proved in \cite{12} that
$\rho(i,\underline{v})$ is always larger for $\gamma_h\not=0$ than
$\gamma_h=0$. Thus, the linear span of the ideal autocorrelation
sequence $\{s_0(t)\}$ is always less than that of other sequences
in the family $\mathcal{F}$.

\hskip 5pt

Run lengths in Eq. (21) deserves further consideration. For $k\geq
2$ and let
\begin{equation}u=1+2^{m}+\cdots+2^{(k-2)m}.\end{equation}
Then
\begin{equation}\delta(i,\underline{v})=\sum\limits_{j=1}^{w(i)}u\cdot2^{mv_j+i_j}
=\sum\limits_{j=1}^{w(i)}\sum\limits_{l=0}^{k-2}2^{m
(v_j+l)+i_j}.\end{equation}

\vspace{2mm}

{\it Lemma 5:} Let $c_{j,l}$ be the remainder of $v_j+l$ modulo
$k$ for $1\leq j\leq w(i)$ and $0\leq l\leq k-2$. Then
\begin{equation}\delta'(i,\underline{v})=\sum\limits_{j=1}^{w(i)}\sum\limits_{l=0}^{k-2}
2^{mc_{j,l}+i_j}.\end{equation}

\vspace{1mm}

{\it Proof:} Since $2^{m(v_j+l)}\equiv
2^{mc_{j,l}}\,\,\mathrm{mod}(2^{mk}-1)$,
$\delta(i,\underline{v})\equiv\delta'(i,\underline{v})\,\mathrm{mod}\,(2^{mk}-1)$.

For a fixed $j$, any two elements of $\{v_j+l\,|\,0\leq l\leq
k-2\}$ are pairwise incongruent modulo $k$. Then
$\{c_{j,l}\,|\,0\leq l\leq k-2\}$ are pairwise different and take
values of $k,k-1,\cdots$, and 1 for a maximal summation. Hence
$$\sum\limits_{j=1}^{w(i)}\sum\limits_{l=0}^{k-2}2^{mc_{j,l}+i_j}
=\sum\limits_{j=1}^{w(i)}2^{i_j}\sum\limits_{l=0}^{k-2}2^{mc_{j,l}}
\leq\sum\limits_{j=1}^{w(i)}2^{i_j}\sum\limits_{l=1}^{k-1}2^{ml}
\leq (2^m-1)\cdot \frac{2^{mk}-2^m}{2^m-1}<2^{mk}-1.$$ Since
$\delta'(i,\underline{v})$ is the remainder of
$\delta(i,\underline{v})$ modulo $2^{mk}-1$, Eq. (25) holds.

\vspace{3mm}

From the proof of Lemma 5 and Eq. (25), the weight of
$\delta'(i,\underline{v})$ is
\begin{equation}{\tau (i,\underline{v})}=(k-1)\cdot
w(i).\end{equation}

\hskip 5pt

To guarantee the period of $\{s_h(t)\}$ reaching $2^n-1$, the
parameter $u$ must be relatively prime to $2^{mk}-1$. The
following lemma gives such an integer.

\vspace{2mm}

{\it Lemma 6:} Let $k\geq 2$ and $u$ be defined as Eq. (23). Then
$${\rm
gcd}(u,\,2^{mk}-1)={\rm gcd}(k-1,\,2^{m}-1).$$

{\it Proof:} Since
$$2^{mk}-1-(2^{2m}-2^m)(1+2^{m}+\cdots+2^{(k-2)m})=2^m-1$$
and $$1+2^{m}+\cdots+2^{(k-2)m}=k-1({\rm mod}\,2^m-1),$$ one has
$${\rm
gcd}(u,\,2^{mk}-1)={\rm gcd}(u,\,2^{m}-1)={\rm
gcd}(k-1,\,2^{m}-1).$$

\hskip 3pt

From this point on we assume ${\rm gcd}(k-1,\,2^{m}-1)=1$. Then
${\rm gcd}(u,\,2^{mk}-1)=1$.

\vspace{3mm}

To simplify Eq. (21), we consider a subfamily of $\mathcal{F}$ as
$$\mathcal{F'}=\{\{s_0(t)\},\{s_h(t)\}: h\not=0,
g_h<\frac{2^{mk}+\varepsilon_h}{2^{m-1}+\varepsilon_h}\,\,{\rm
and}\,\,0\leq c_h\leq 2^{mk-1}\},$$ and estimate a lower bound for
linear spans of sequences in this subfamily. This subfamily
contains a great majority of the sequences in $\mathcal{F}$ as
shown by
$$|\mathcal{F'}|>\frac{1}{2}(\phi(2^{mk}-1)+\phi(2^{mk}+1)),$$\cite{13}, where
$\phi(t)$ is Euler's phi function. The subfamily size is close to
$2^{mk}$.

\vspace{2mm}

For a sequence $\{s_h(t)\}$ in $\mathcal{F'}$ with $h\not=0$, we
have
\begin{equation}\rho(i,\underline{v})=\prod\limits^{R(i,\underline{v})}_{j=1}\{
2^{L(i,\underline{v},\,j)+1}-1\}\end{equation} for any $i\in I$
and $\underline{v}\in V^{w(i)}$.  We use an approach proposed by
Klapper \cite{13} to estimate a lower bound on
$\sum\limits_{\underline{v}\in V^{w(i)}}\rho(i,\underline{v})$ for
some $i$.

\vspace{3mm}

For $1\leq t\leq m-1$, let $i^{(t)}=\sum\limits_{j=1}^{t}2^{j-1}$
with the weight $t$.

\vspace{3mm}

{\it Lemma 7:} Let $1\leq t\leq m-1$. Then

(1) For $\gamma_h=0$,
$$\sum\limits_{\underline{v}\in V^{t}}\rho
(i^{(t)},\underline{v})=(2^{k-1}k)^{t}.$$

(2) For $\gamma_h\neq 0$,
$$\sum\limits_{\underline{v}\in V^{t}}\rho
(i^{(t)},\underline{v})>3^{k-1}k((3k-1) 2^{k-2})^{t-1}.$$

{\it Proof:} (1) The conclusion follows that for each
$\underline{v}\in V^t$, $\rho (i^{(t)},\underline{v})=2^{(k-1)t}$
by Eq. (22) and Eq. (26).

(2) Assume $\gamma_h\neq 0$. We establish a lower bound on
$\sum\limits_{\underline{v'}\in V^{t+1}}\rho
(i^{(t+1)},\underline{v'})/\sum\limits_{\underline{v}\in
V^{t}}\rho (i^{(t)},\underline{v})$ for $1\leq t\leq m-2$ and then
deduce the conclusion.

For any $\underline{v}=(v_1,\cdots,v_t)\in V^{t}$ and $v_{t+1}\in
V$, let $\underline{v'}=(v_1,\cdots,v_t,v_{t+1})\in V^{t+1}$. By
Eq. (24) and Eq. (25),
$$\delta(i^{(t)},\underline{v})=\sum\limits_{j=1}
^{t}\sum\limits_{l=0}^{k-2}2^{m(v_j+l)+j-1}\,\,{\rm and}\,\,\,
\delta'(i^{(t)},\underline{v})=\sum\limits_{j=1}
^{t}\sum\limits_{l=0}^{k-2}2^{mc_{j,l}+j-1}.$$ There are similar
expressions for $\delta(i^{(t+1)},\underline{v'})$ and
$\delta'(i^{(t+1)},\underline{v'})$. Define
$$\widetilde{\delta}(i^{(t)},\underline{v})=\sum\limits_{j=1}
^{t}\sum\limits_{l=0}^{k-1}2^{mc_{j,l}+j-1}.$$

For fixed integers $d$ and $j$ ($0\leq d\leq k-1$ and $1\leq j\leq
t$), there exists a unique integer $l$ with $0\leq l\leq k-1$ such
that $c_{j,l}=d$. This indicates that all run intervals of
$\widetilde{\delta}(i^{(t)},\underline{v})$ are
\begin{equation}
[0,t-1],[m,m+t-1],\cdots,[m(k-1),m(k-1)+t-1].
\end{equation}
Similarly, the run intervals of
$\widetilde{\delta}(i^{(t+1)},\underline{v'})$ are
$$
[0,t],[m,m+t],\cdots,[m(k-1),m(k-1)+t].
$$
By deleting all terms with the form $2^{mc_{j,k-1}+j-1}$ ($1\leq
j\leq t$) from the binary expansion of
$\widetilde{\delta}(i^{(t)},\underline{v})$, the binary expansion
of $\delta'(i^{(t)},\underline{v})$ is obtained. Thus, the run
intervals of $\delta'(i^{(t)},\underline{v})$ can be obtained by
deleting the integers $mc_{j,k-1}+j-1$ ($1\leq j\leq t$) from the
run intervals in Eq. (20).

A run interval of $\delta'(i^{(t)},\underline{v})$ is called a
type-I interval if it contains an integer of form $mc_{t,l}+t-1$,
where $0\leq l\leq k-2$, and is called a type-II interval
otherwise. Thus, $\delta'(i^{(t)},\underline{v})$ has exactly
$(k-1)$ run intervals in type-I. Let $u_l$ denote the length of
the run interval containing $mc_{t,l}+t-1$.

\vspace{2mm}

When $v_{t+1}=v_t$, for any $0\leq l\leq k-1$, one has
$$v_{t+1}+l=v_t+l\,\,{\rm and}\,\,mc_{t+1,l}+t=(mc_{t,l}+t-1)+1.$$
This means that the length of each type-I run interval of
$\delta'(i^{(t+1)},\underline{v'})$ is larger by 1 than that of a
corresponding type-I run interval of
$\delta'(i^{(t)},\underline{v})$, and that all type-II run
intervals of $\delta'(i^{(t+1)},\underline{v'})$ coincide with
that of $\delta'(i^{(t)},\underline{v})$. (Example 8 (1)
illustrates this.) Thus,
\begin{equation}
\frac{\rho (i^{(t+1)},\underline{v'})}{\rho
(i^{(t)},\underline{v})} =\prod\limits_{l=0}^{k-2}
\frac{2^{u_l+1+1}-1}{2^{u_l+1}-1}
>\prod\limits_{l=0}^{k-2}2=2^{k-1}.
\end{equation}

When $v_{t+1}\neq v_{t}$, one has
$$mc_{t+1,l'}+t=(mc_{t,l}+t-1)+1$$
if and only if
\begin{equation}l'=l+v_t-v_{t+1}(\mathrm{mod}\,k).\end{equation}
Let $l_0$ $(0\leq l_0\leq k-1)$ be the unique solution of
$$l+v_t-v_{t+1}=k-1(\mathrm{mod}\,k).$$
Then $0\leq l_0\leq k-2$.

For any $0\leq l\leq k-2$ with $l\neq l_0$, let $0\leq l'\leq k-2$
be determined by Eq. (29). Then among the run intervals of
$\delta'(i^{(t+1)},\underline{v'})$, the length of the interval
containing the integer $mc_{t+1,l'}+t$ is larger by 1 than that of
the interval of $\delta'(i^{(t)},\underline{v})$ containing
$mc_{t,l}+t-1$. On the other hand, the interval of
$\delta'(i^{(t)},\underline{v})$ containing the integer
$mc_{t,l_0}+t-1$ is identical to a corresponding interval of
$\delta'(i^{(t+1)},\underline{v'})$. So does each type-II interval
of $\delta'(i^{(t)},\underline{v})$. Notice that the integer
$mc_{t,k-1}+t$ is not in any interval of
$\delta'(i^{(t)},\underline{v})$, and
$[mc_{t,k-1}+t]=[mc_{t+1,l_1}+t]$ is a single-point run interval
of $\delta'(i^{(t+1)},\underline{v'})$, where
$l_1=k-1+v_t-v_{t+1}(\mathrm{mod}k)$ and $0\leq l_1\leq k-2$. (
Example 8 (2) illustrates this .) Thus,
\begin{equation}
\frac{\rho (i^{(t+1)},\underline{v'})}{\rho
(i^{(t)},\underline{v})} =(2^{1+1}-1)\prod\limits_{l=0,l\neq
l_0}^{k-2} \frac{2^{u_l+1+1}-1}{2^{u_l+1}-1}
>3\cdot 2^{k-2}.\end{equation}

Applying Eq. (29) and Eq. (31), one has
\begin{equation}\begin{array}{ll}\sum\limits_{\underline{v'}\in V^{t+1}}\rho
(i^{(t+1)},\underline{v'})
 &=\sum\limits_{\underline{v}\in
V^{t}}(\sum\limits_{v_{t+1}=v_t}\rho
(i^{(t+1)},(\underline{v},v_{t+1}))+\sum\limits_{v_{t+1}\not=v_t}\rho
(i^{(t+1)},(\underline{v},v_{t+1}))\\
&>\sum\limits_{\underline{v}\in V^{t}}(2^{k-1}+(k-1)3\cdot
2^{k-2})\rho (i^{(t)},\underline{v})\\
&=(3k-1)\cdot 2^{k-2}\sum\limits_{\underline{v}\in V^{t}}\rho
(i^{(t)},\underline{v}).\end{array}\end{equation}

For $v_1\in V=\{0,1,\cdots,k-1\}$, one has
$\delta(1,v_1)=\sum\limits_{l=0}^{k-2}2^{m(v_1+l)}$ and
$\delta'(1,v_1)=\sum\limits_{l=0}^{k-2}2^{mc_{1,l}}$. There are
exactly $(k-1)$ 1-runs of length 1. Thus,
$$\rho(i^{(1)},v_1)=\rho
(1,v_1)=\prod\limits_{l=0}^{k-2}(2^{1+1}-1)=3^{k-1},$$ and
\begin{equation}\sum\limits_{v_1\in V}\rho (1,v_1)=k\cdot
3^{k-1}.\end{equation} Applying Eq. (33), and Eq. (32)
iteratively, one has Lemma 7 (2).

\vspace{3mm}

{\it Example 8:} (1) Suppose that $m=7$, $k=t=4$,
$\underline{v}=(3,0,3,1)$ and $\underline{v'}=(3,0,3,1,1)$. The
run intervals of $\widetilde{\delta}(i^{(5)},\underline{v'})$ and
$\widetilde{\delta}(i^{(4)},\underline{v})$ are
$$[0,3],[7,10],[14,17],[21,24]$$ and
$$[0,4],[7,11],[14,18],[21,25],$$
respectively. A direct calculation will find the run intervals of
$\delta(i^{(4)},\underline{v})$ and
$\delta(i^{(5)},\underline{v'})$ are
$$[0,2],[7,10]^{*},[15],[17]^{*},[21],[23,24]^{*}$$
and
$$[0,2],[7,11]^{*},[15],[17,18]^{*},[21],[23,25]^{*},$$
respectively, where the intervals marked with $*$ are in type-I
and type-II otherwise.

Obviously, the type-I run intervals $[7,11],[17,18]$, and
$[23,25]$ are of lengths larger by 1 than $[7,10],[17]$, and
$[23,24]$, respectively, and all type-II run intervals of
$\delta(i^{(5)},\underline{v'})$ and
$\delta(i^{(4)},\underline{v})$ coincide.

\vspace{2mm}

(2) If $\underline{v'}=(3,0,3,1,2)$, then the run intervals of
$\delta(i^{(5)},\underline{v'})$ are
$$[0,2],[4],[7,10]^{+},[15],[17,18]^{*},[21],[23,25]^{*}.$$
Since $l_0=k-1+v_{t+1}-v_t=0\,({\rm mod}\, k)$, for $0\leq l\leq
k-2$ with $l\neq l_0$, i.e., for $l=1$ or 2, $l'=l+v_t-v_{t+1}=0$
or $1$. Then
$$\{mc_{t+1,l'}+t\,|\,l'=0,2\}=\{18,25\},$$
and we get two type-I run intervals marked with $*$, i.e.,
$[17,18]$ and $[23,25]$. Since $l_1=k-1+v_t-v_{t+1}=2({\rm
mod}\,k)$, the remaining type-I run interval is the single-point
set $[4]$. The type-I interval of $\delta(i^{(4)},\underline{v})$
containing $mc_{t,l_0}+t-1=10$ is [7,10], it is a type-II interval
of $\delta(i^{(5)},\underline{v'})$, which is marked with $+$.
Other type-II run intervals of $\delta(i^{(5)},\underline{v'})$
and $\delta(i^{(4)},\underline{v})$ coincide.

\vspace{2mm}

Now we deduce the main result of the correspondence as follows.

By the assumption, we have $i^{(m-1)}\in I\cap \Gamma(m)$. The
size of the cyclotomic coset containing $i^{(m-1)}$ is $m$.
Applying Proposition 4 to such an index set $I$ gives
\begin{equation}LS(\{s_h(t)\})\geq m\cdot \sum\limits_{\underline{v}\in
V^{m-1}}\rho (i^{(m-1)},\underline{v}).\end{equation} Applying
Lemma 7 to Eq. (32), one has the theorem below.

\vspace{2mm}

{\it Theorem 9:} Let $\{s_h(t)\}\in \mathcal{F'}$.

(1)
$$LS(\{s_0(t)\})\geq L_0=m(2^{k-1}k)^{m-1}.$$

(2) For $h\not=0$,$$LS(\{s_h(t)\})>L_1=
3^{k-1}mk[2^{k-2}(3k-1)]^{m-2}.$$

\vspace{2mm}

For a large integer $n$, the lower bound $L_1$ given in Theorem 9
is maximized when $k=4$. By Lemma 6, we choose $k=4$ when $m$ is
odd and choose $k=3$ or 5 when $m$ is even. Table II lists the
bounds $L_0$ and $L_1$.

\begin{table}\caption{The lower bound of linear span of sequences with period $2^n-1$ in
family $\mathcal{F'}$}
\begin{center}
  \begin{tabular*}{0.78\textwidth}
     {@{\extracolsep{\fill}}|c|c|c|c|} \hline
  $k$ &$3$\,\, &$4$\,\,&$5$ \\[0.7ex]
  \hline
  $n$ &$6m$\,\, &$8m$\,\,&$10m$ \\[0.7ex]
  \hline
 $L_0$ & $12^{\frac{n}{6}}n/72\,\,$&
 $2^{\frac{5n}{8}}n/256$
 &$80^{\frac{n}{10}}n/800$\\[0.7ex]
    \hline
 $L_1$ & $9n\cdot 2^{\frac{2n}{3}}/512$&
 $27n\cdot 44^{\frac{n}{8}}/3872$
 &$81n\cdot 112^{\frac{n}{10}}/25088$\\[0.7ex]
\hline
    \end{tabular*}
  \end{center}
\end{table}

\vspace{2mm}

{\it Remark 10: } The bounds $O(n\cdot 2 ^{\frac{2n}{3}})$,
$O(n\cdot 44^{\frac{n}{8}})$ and $O(n\cdot 112^{\frac{n}{10}})$,
given by taking $k=3,4,5$, respectively, are exponentially larger
than that of No sequences and TN sequences, whose bounds are
$O(n\cdot 4^{\frac{n}{4}})$ and $O(n\cdot 5^{\frac{n}{4}})$,
respectively \cite{13}. If we take $k=2$, the lower bounds in
Theorem 9 will be the same as that of TN sequences.

More precisely, let $U_{No}=2^{\frac{n}{2}}\cdot n/2$ and
$U_{TN}=9n\cdot(16/3)^{\frac{n}{4}-3}$. Then $U_{No}$ and $U_{TN}$
are upper bounds on linear spans of No sequences and TN sequences,
respectively \cite{12,13}, which is exponentially smaller than the
lower bounds in Theorem 9 (2), since
$44^{\frac{1}{8}}>112^{\frac{1}{10}}>2^{\frac{2}{3}}>(16/3)^{\frac{1}{4}}>2^{\frac{1}{2}}$.

\section{An extension to sequences with ideal autocorrelation}

Instantiating the ideal autocorrelation sequence in Eq. (6), we
can tighten the bound in Theorem 9 (1), and construct a class of
ideal autocorrelation sequences with larger linear span. Most of
known ideal autocorrelation sequences have very small linear span
\cite{21}, \cite{22}, \cite{23}. Legendre sequences of a prime
period can achieve an upper bound on linear span of binary ideal
autocorrelation sequences \cite{18}.

\vspace{2mm}

Let $p=2^m-1$ be a Mersenne prime for some prime $m\geq 3$. A
Legendre sequence of period $p$ is defined as $\{a(t)\}$ where
$$
a(t)=\left\{\begin{array}{ll}1,&{\rm if}\,\, t\equiv 0\, ({\rm
mod}p); \\0,&{\rm if}\,\,t\,\, \mathrm{is \,\,a\,\, quadratic\,\,
residue\,\, modulo}\,\, p;
\\1,&{\rm if}\,\, t\,\,\mathrm{is\,\, a \,\,quadratic\,\, nonresidue\,\, modulo}\,\,
p.
\end{array}\right.
$$
It is easy to verify that $\{a(t)\}$ is an ideal autocorrelation
sequence. Furthermore, its trace representation is given as
follows.

\vspace{2mm}

{\it Lemma 11:} (Main theorem of \cite{19}) Let $\gamma$ be a
primitive element of $Z_p$. There is a primitive element $\beta$
of $F_{2^m}$ such that
$$
a(t)=\sum\limits^{\frac{p-1}{2m}-1}_{j=0}tr^{m}_{1}(\beta^{\gamma^{2j}t})$$
is the trace representation of $\{a(t)\}$.

\vspace{2mm}

For $\zeta=0$ or $1$, define two sequences $\{a^{(\zeta)}(t)\}$
where
$$
a^{(\zeta)}(t)=\sum\limits^{\frac{p-1}{2m}-1}_{j=0}tr^{m}_{1}(\beta^{t\gamma^{2j+\zeta}}).$$
Then, $\{a^{(0)}(t)\}=\{a(t)\}$, and $\{a^{(1)}(t)\}$ is the
$\gamma$-decimation of $\{a(t)\}$. Therefore, both sequences have
ideal autocorrelation property.

\vspace{2mm}

Let $k\geq 2$ and $u=1+2^{m}+\cdots+2^{(k-2)m}$. Assume ${\rm
gcd}(k-1,p)=1$. We construct ideal autocorrelation sequences from
$\{a^{(0)}(t)\}$ and $\{a^{(1)}(t)\}$ as follows. For $\zeta=0$ or
$1$, define
\begin{equation} s^{(\zeta)}(t)=\sum\limits^{\frac{p-1}{2m}-1}_{j=0}
tr_1^m(\{tr^{mk}_{m}[(tr^{n}_{mk}(\alpha^{2t}))^u]\}^{\gamma^{2j+\zeta}}).\end{equation}
By Lemma 1, $\{s^{\zeta}(t)\}$ is an ideal autocorrelation
sequence of period $2^{2mk}-1$.

\vspace{2mm}

The following lemma is needed for deducing a tighter bound on the
linear span of $\{s^{(\zeta)}(t)\}$.

\vspace{2mm}

{\it Lemma 12:} (1) (\cite{19}) When $i$ varies from $0$ to
$\frac{p-1}{m}-1$, $\gamma^i$ runs through all the $\frac{p-1}{m}$
cyclotomic cosets of size $m$ modulo $p$. For some integer $j$,
$\gamma^{\frac{p-1}{m}}=2^j$.

(2) Among $\frac{p-1}{m}$ cyclotomic cosets of size $m$ modulo
$p$, the number of cosets consisting of integers of weight $i$ is
$\left(m\atop{i}\right)/m$.

\vspace{2mm}

{\it Theorem 13:} For either $\zeta=0$ or $1$, the linear span of
sequences defined as in Eq. (33) satisfies
$$LS(\{s^{(\zeta)}(t)\})\geq\frac{1}{2}[(1+2^{k-1}k)^m-1-(2^{k-1}k)^m].$$

{\it Proof:} For $\zeta=0$ or $1$, Proposition 4 together with Eq.
(22) and Eq. (26) yields
$$LS(\{s^{(\zeta)}(t)\})=\sum\limits^{\frac{p-1}{2m}-1}_{j=0}m\cdot
(2^{k-1}k)^{w(\gamma^{2j+\zeta})}.$$ One has
$$\begin{array}{rcl}&&LS(\{s^{(0)}(t)\})+LS(\{s^{(1)}(t)\})\\&=&\sum\limits^{\frac{p-1}{2m}-1}_{j=0}m\cdot
(2^{k-1}k)^{w(\gamma^{2j})}+\sum\limits^{\frac{p-1}{2m}-1}_{j=0}m\cdot
(2^{k-1}k)^{w(\gamma^{2j+1})}\\
&=&\sum\limits^{\frac{p-1}{m}-1}_{j=0}m\cdot
(2^{k-1}k)^{w(\gamma^{j})}\\
&=&\left(m\atop 1\right)\cdot (2^{k-1}k)+\left(m\atop
2\right)\cdot (2^{k-1}k)^2+\cdots+\left(m\atop{ m-1}\right)\cdot
(2^{k-1}k)^{m-1}\\&=&(1+2^{k-1}k)^m-1-(2^{k-1}k)^m.\end{array}$$
Thus, Theorem 13 holds.

 \vspace{3mm}

{\it Remark 14:} An analysis to $L_0=m(2^{k-1}k)^{m-1}$ show that,
for any given $n$, the bound $L_0$ is maximized only if $k\leq 6$.
In this case, if $m\geq 2^kk+1$, then the bound in Theorem 13 is
tighter than that in Theorem 9 (1). More precisely,
$$\frac{1}{2}[(1+2^{k-1}k)^m-1-(2^{k-1}k)^m]\geq m(2^{k-1}k)^{m-1}$$
holds for $k\leq 6$ and $m\geq 2^kk+1$.

\vspace{3mm}

Sequences defined in Eq. (33) with the period of $2^{2mk}-1$ are
an application of the construction of Eq. (7) to the case of
$k\geq 2$. If we take $k=1$ and define
\begin{equation} \widetilde{s}^{(\zeta)}(t)=\sum\limits^{\frac{p-1}{2m}-1}_{j=0}
tr_1^m([tr^{2m}_{m}(\alpha^{2t})]^{\gamma^{2j+\zeta}}),\end{equation}
($\zeta=0$ or 1), we will get two ideal autocorrelation sequences
of period $2^{2m}-1$, and their linear span can be shown as
$$LS(\{\widetilde{s}^{(\zeta)}(t)\})=\sum\limits^{\frac{p-1}{2m}-1}_{j=0}m\cdot
2^{w(\gamma^{2j+\zeta})}$$ by Proposition 4 and Eq. (22). An
analysis similar to Theorem 13 shows either
$\{\widetilde{s}^{(0)}(t)\}$ or $\{\widetilde{s}^{(1)}(t)\}$ has
linear span not less than $(3^m-1-2^m)/2$.

\vspace{2mm}

Eq. (33) and Eq. (34) provide a way to generate ideal
autocorrelation sequences with large linear span.

\vspace{2mm}

{\it Example 15:} Let
$\{a(t)=\sum\limits^{8}_{j=0}tr^{7}_{1}(\alpha^{3^{2j}t})\}$ be a
Legendre sequence of period 127 and $\{b(t)=a(3t)\}$ be its
$3$-decimation. The linear span of the sequence $\{s^{(1)}(t)\}$
derived from $\{b(t)\}$ is $1232>1029=(3^7-1-2^7)/2$, which is
larger than that of the sequence of period $2^{14}-1$ given in
Example 9 of \cite{23}.

\section{Concluding remarks}

The generalized Kasami sequence set \cite{10} is given by
$$\Gamma=\{g(tr^n_{n/2}(x^2)+\beta x^{2^{mk}+1}),\beta\in F_{2^{mk}},x\in F^{*}_{2^n}\},$$
where $\{g(x),x\in F^{*}_{2^{mk}}\}$ is any one sequence with
ideal autocorrelation property. The set $\Gamma$ has optimal
correlation property with respect to Welch bound.   The linear
span of sequences in $\Gamma$ depends on $g(x)$.

Let $g(x)=\sum_{i\in I}[tr^{mk}_m(x^u)]^i$.   Consider the linear
span of sequences in $\Gamma$. To obtain large linear span, an
efficient approach is to choose $u$ and index set $I$
appropriately such that  the integer $ \delta'(i,\underline{v})$
has large binary weight,
$$
\delta'(i,\underline{v})=\sum\limits_{j=1}^{w(i)}u\cdot2^{mv_j+i_j}\,({\rm
mod \, 2^{mk}-1}),$$ where $i\in I$ and
$\underline{v}=(v_1,v_2,\cdots,v_{w(i)})\in V^{w(i)}$. In the
original Kasami construction and No sequences, $I$ was equal to
$\{1\}$.  No sequences achieved large linear span by having $u$
with large weight.  Klapper took $u=1$ and $I$ consisting of only
one integer with large binary weight, such that TN sequences can
obtain even larger linear span than No sequences and the small set
of Kasami sequences.

This correspondence discusses a new  case where both $u$ and one
element in $I$ have large binary weight.  For appropriate
parameters $(m,k,u,I)$, sequences discussed in this correspondence
can obtain larger linear span than that of Kasami sequences (small
set), No sequences and  TN sequences.

Very likely by choosing $u$ with other forms, sequences with
larger linear span can be found.


%
%
\appendices

\end{document}